\documentclass[11pt,letterpaper]{article}
\pdfoutput=1
\usepackage{jheppub}
\usepackage{color}
\usepackage{graphicx}

\usepackage{comment}
\usepackage{verbatim}
\usepackage{amsmath}
\usepackage{amssymb}
\usepackage{subfigure}
\usepackage{url}
\usepackage{slashed}
\usepackage{multirow}
\usepackage{threeparttable}
\usepackage{paralist}
\usepackage{comment}

\usepackage{accents}
\newlength{\dhatheight}
\newcommand{\doublehat}[1]{
    \settoheight{\dhatheight}{\ensuremath{\hat{#1}}}
    \addtolength{\dhatheight}{-0.35ex}
    \hat{\vphantom{\rule{1pt}{\dhatheight}}
    \smash{\hat{#1}}}}

\usepackage{color}
\definecolor{darkgreen}{rgb}{0,0.5,0}

\DeclareRobustCommand{\Sec}[1]{Sec.~\ref{#1}}

\DeclareRobustCommand{\Tab}[1]{Table~\ref{#1}}

\DeclareRobustCommand{\Fig}[1]{Fig.~\ref{#1}}

\DeclareRobustCommand{\Eq}[1]{Eq.~(\ref{#1})}

\newcommand{\be}{\begin{equation}}
\newcommand{\ee}{\end{equation}}

\begin{document}

\title{The Social Higgs}

\author{Daniele Bertolini and}
\emailAdd{danbert@mit.edu}

\author{Matthew McCullough}
\emailAdd{mccull@mit.edu}
\affiliation{Center for Theoretical Physics, Massachusetts Institute of Technology, Cambridge, MA 02139, USA}

\date{\today}

\abstract{Using published Higgs search data we investigate whether any evidence supports the possibility that the Higgs may be mixed with other neutral scalars.  We combine the positive evidence for the Higgs at $125.5$ GeV with search constraints at other masses to explore the viability of two simple models.  The first Higgs `friend' model is simply a neutral scalar mixed with the Higgs.  In the second Higgs `accomplice' model the new scalar has an enhanced coupling to photons due to couplings to additional charged fields.  We find that the latter scenario allows improvement in fitting the data by accommodating enhanced $h\rightarrow \gamma \gamma$ rates and suppression in other channels for a Higgs mass of $125.5$ GeV.  Small excesses at other masses allow the additional scalar to further improve the fit to the data, particularly if it has mass in the vicinity of $210$ GeV.  Due to observed event rates at $125.5$ GeV and strong limits in high mass Higgs searches, mixing angles $\theta\gtrsim\pi/4$ are typically disfavored at the $95\%$ confidence level, depending on the mass of the scalar.}

\preprint{MIT-CTP {4383}}

\maketitle

\section{Introduction}
\label{sec:introduction}
The recent announcement of significant evidence for the Higgs boson \cite{CMSan,ATLASan,:2012gu,:2012gk}, gathered at the CERN LHC, brings particle physics into a new era of discovery.  Further data may confirm this signal to be due to the Standard Model (SM) Higgs, however there are already emerging, albeit very weak, hints that production cross-sections and/or decay rates might not be quite as expected for the SM Higgs.  If these hints persist then it may be the case that the discovery of the Higgs comes accompanied by convincing evidence for physics beyond the SM (BSM physics).  From a theoretical perspective some modification of Higgs physics has long been expected, since substantial theoretical motivation for BSM physics is aimed at resolving the hierarchy problem, which is concerned with the Higgs sector.  Solutions to this problem often require the existence of additional electroweak-charged states and/or additional scalars coupled to, or mixed with, the Higgs.

Even if one abandons the hierarchy problem as motivation there is always the possibility that additional hidden sectors exist, perhaps related to dark matter.  The Higgs sector of the SM contains a super-renormalizable Lorentz and gauge invariant operator, which can easily accommodate couplings to new hidden sector physics, the so-called `Higgs Portal' \cite{Silveira:1985rk,McDonald:1993ex,Burgess:2000yq,Davoudiasl:2004be,Patt:2006fw,MarchRussell:2008yu,Andreas:2010dz,Raidal:2011xk,He:2011de,Mambrini:2011ik,Englert:2011yb,Batell:2011pz,Djouadi:2012zc,Mambrini:2011ri}.  Such couplings may allow for Higgs decays to neutral particles, leading to an additional invisible width for the Higgs.  The main consequence of this scenario is that all detectable branching ratios become equally suppressed, leading to a democratic reduction in the Higgs signal rates.  In addition to this, hidden sector scalars can also mix with the Higgs through the Higgs portal interaction.

Motivated by these simple considerations we study the implications of the LHC Higgs searches on simple models of a singlet scalar mixed with the Higgs.  Although simplified, we believe these models should map on to some theoretically motivated scenarios, such as the Next to Minimal Supersymmetric Standard Model (NMSSM) where the Higgs mixes with an extra singlet which has and induced coupling to photons through its Yukawa coupling to charged Higgsinos.

If additional neutral scalars mix with the Higgs the mass eigenstates and interaction eigenstates are not aligned and the properties of the Higgs are altered, similar to the so-called `Higgs look-alike', or `Higgs Friend' scenario \cite{Hubisz:2008gg,DeRujula:2010ys,DeRujula:2010kt,Fox:2011qc,Gupta:2011gd}.\footnote{Although we do not assume that the neutral scalars have couplings to extra colored particles, as in \cite{Fox:2011qc}, the scenario considered here is sufficiently similar in spirit to the proposals in \cite{Fox:2011qc} that we adopt the `Higgs Friend' terminology.}  If this mixing alone is present the individual branching ratios of the Higgs remain the same and the production of the Higgs is suppressed.  The overall effect is thus, at the level of the current Higgs searches, indistinguishable from the case where the Higgs has an additional invisible width.  However, with careful study the two can be distinguished.  For example, an invisible width can be measured by searching for mono-jet signals coming from initial state radiation in Higgs production \cite{Ellis:1987xu,Djouadi:1991tka,Dawson:1990zj,Dawson:1991au,Spira:1995rr,Djouadi:2012zc}.  On the other hand, mixing with a neutral scalar can be confirmed more directly by searching for the extra scalar, exploiting the fact that it inherits many properties of the Higgs.  In this work we are concerned with the latter scenario.  We assume that the Higgs is present, with mass of $125.5$ GeV, and then consider limits from the Higgs searches on a Higgs friend.  Of course, it is plausible that the friend might be much more massive and thus effectively decoupled, in which case it would be very difficult to unambiguously confirm its presence.  We consider fits to this scenario as well.

As stated, this simple Higgs friend scenario leads to suppression of all Higgs signal rates, independent of the particular search channel under consideration.  However, early evidence from the LHC suggests that while some small degree of suppression in most channels may provide a better fit to the data, the $h\rightarrow \gamma \gamma$ channel, which is driving the statistical significance of the discovery, is possibly enhanced to some degree.  As such, it is clear that the Higgs friend scenario alone will not lead to significant improvement in fitting the data when compared to the SM.  For this reason we consider the addition of charged vector-like fields, which could be scalar or fermionic, and couple to the Higgs friend.\footnote{It is often the case that additional electroweak-charged fields are present in extensions of the SM Higgs sector, so the introduction of extra charged fields is a plausible augmentation of the Higgs friend scenario.}  This interaction enhances the coupling of the friend to photons at one-loop.  Once the Higgs mixes with the friend this can enhance the Higgs decays to photons, allowing a better fit to the data than the SM Higgs.  We call this the `Higgs accomplice' scenario.  

In order to test the viability of the Higgs friend and Higgs accomplice scenarios it is necessary to confront these models with data.  Since the first convincing hints of a Higgs were reported in December 2011 there has been significant interest in determining how best to extract Higgs couplings from the data, leading to a number of studies \cite{Azatov:2012bz,Carmi:2012yp,Espinosa:2012ir,Giardino:2012ww,Azatov:2012rd,Low:2012rj,Corbett:2012dm,Giardino:2012dp,Buck,Ellis,Riva,Groj,Falk}.  We do this by approximating the likelihood functions for the Higgs signal strength in particular decay channels by using the best fit values or, in one case, the expected versus observed $95\%$ exclusion contours provided by the collaborations.  This information can then be used to estimate best fit points and confidence contours for the models considered.

The finite mass-resolution of the Higgs searches introduces additional subtleties when considering the presence of two Higgs-like scalars contributing signal events in the searches.  Whenever the mass separation of the Higgs and friend is much greater than the mass resolution of the searches we can take the product of the individual likelihoods, since the searches at different masses are approximately independent.  However, when they are close enough in mass that signal events from both scalars cannot be considered separately a more sophisticated likelihood must be constructed.  Without performing a collider simulation we suggest a crude method by which to construct the combined likelihood, which we believe captures the dominant features of the likelihood function for the Higgs and the extra scalar in this case.  Rather than using this combined likelihood to make precise statements about scenarios where both scalars are close in mass, we instead use it to estimate the mass range in which the searches are effectively independent, to determine when the combined likelihood can be trusted.  We then find the best fit parameters and associated $95\%$ confidence contours for both scenarios.

Before presenting our results in \Sec{sec:results} we will briefly review the Higgs friend and Higgs accomplice scenarios in \Sec{sec:friendliness} and our statistical methods in \Sec{sec:stats}.  We also consider precision electroweak constraints in \Sec{sec:electroweak} and draw conclusions in \Sec{sec:discussion}. 

\section{Higgs Friends and Accomplices}
\label{sec:friendliness}
We consider a simple set-up in which an extra field, $s$, mixes with the neutral Higgs through a Higgs portal coupling.  In the mass-eigenstate basis the two neutral scalars are $\tilde{h}$ and $\tilde{s}$, which are related to the interaction eigenstates through
\be
\left( \begin{array}{c}
h \\
s \end{array} \right) = \left( \begin{array}{cc}
\cos{\theta} & -\sin{\theta} \\
\sin{\theta} & \cos{\theta} \end{array} \right) \cdot \left( \begin{array}{c}
\tilde{h} \\
\tilde{s} \end{array} \right).
\ee
This is the Higgs friend scenario.  We also consider the Higgs accomplice scenario in which $s$ couples to additional charged particles.  At one loop this leads to a coupling
\be
\mathcal{L} = \alpha c_{h \gamma \gamma} s F^{\mu \nu} F_{\mu \nu} ~~,
\ee
where $c_{h \gamma \gamma}$ is the usual SM coupling of the Higgs to photons, and $\alpha$ parameterizes deviations from this coupling.\footnote{In general $\alpha$ can either be positive or negative.}  Without loss of generality we impose $0<\theta < \pi/2$.  We will refer to the scalar at $125.5$ GeV, $\tilde{h}$, as the Higgs.

All relevant Higgs production cross-sections at the LHC now come suppressed by a factor of $\cos^2{\theta}$ and every decay width is suppressed by the same factor, with the exception of decays to photons which are now accompanied by a factor of $(\cos{\theta}+\alpha \sin{\theta})^2$.  Since decays to photons are far subdominant then, to a good approximation, all branching ratios remain the same as the SM Higgs, with the exception of the branching ratio to photons which is accompanied by the factor $(1+\alpha \tan{\theta})^2$.  Thus for the search channels $h \rightarrow bb, \tau \tau, WW, ZZ$ the total event rate normalized to the event rate for a SM Higgs, otherwise known as the strength modifier, $\mu$, is simply $\mu = \cos^2{\theta}$ and for $h \rightarrow \gamma \gamma$ it is $\mu_{\gamma \gamma} = (\cos{\theta}+\alpha \sin{\theta})^2$.

Production of the Higgs friend, $\tilde{s}$, is suppressed by a factor of $\sin^2{\theta}$ compared to SM Higgs production.  Whenever $m_{\tilde{s}} < 2 m_{\tilde{h}}$, the strength modifier for the friend in the diphoton channel is $(\sin{\theta}-\alpha \cos{\theta})^2$ and is $\sin^2{\theta}$ for all other channels.

Whenever $m_{\tilde{s}} > 2 m_{\tilde{h}}$ the trilinear scalar interactions allow for the decays $\tilde{s} \rightarrow 2 \tilde{h}$.  These decays could lead to interesting signatures, such as $4b$ final states, however such signals are not currently accessible at the LHC.\footnote{We thank Christoph Englert for discussions on this point.}  As we are considering the sensitivity of the dedicated Higgs searches to Higgs friends and accomplices we can treat this additional width as invisible.  Given the physical masses, mixing angle, and scalar potential parameters, one can determine the magnitude of this interaction, which is essentially a free parameter, and the resultant width (see e.g.\ \cite{Englert:2011yb}).  As the trilinear coupling is a free parameter we do not lose generality by taking the invisible branching ratio as a free parameter.\footnote{Both $\tilde{h}$ and $\tilde{s}$ could also have additional widths to invisible states.  For $\tilde{s}$ this is automatically accommodated in this analysis since the invisible width is a free parameter.  For $\tilde{h}$ the overall effect is to democratically reduce event rates.}  We can express this branching ratio in a model-independent sense as   
\be
\text{BR}(\tilde{s} \rightarrow 2 \tilde{h}) = \kappa \left( 1 - 4 \frac{m^2_{\tilde{h}}}{m^2_{\tilde{s}}} \right)^{1/2}  ~~,
\label{eq:BR}
\ee
where $\kappa$ is the branching ratio in the limit $m_{\tilde{s}}\rightarrow \infty$ and the kinematic factors are included such that the branching ratio vanishes at threshold.

Hence for a Higgs friend or accomplice the strength modifiers become
\be
\mu = \sin^2{\theta}  \left(1-\kappa \sqrt{ 1 - 4 \frac{m^2_{\tilde{h}}}{m^2_{\tilde{s}}} } \right) ~~, ~~ \mu_{\gamma \gamma} = (\sin{\theta}-\alpha \cos{\theta})^2 \left(1-\kappa \sqrt{ 1 - 4 \frac{m^2_{\tilde{h}}}{m^2_{\tilde{s}}} } \right) ~~,
\ee
for the $\tilde{s} \rightarrow bb, \tau \tau, WW, ZZ$ and $\tilde{s} \rightarrow \gamma \gamma$ channels respectively.

Counting parameters, we have the recently determined parameter of the SM, $m_{\tilde{h}}$, along with four new free parameters $m_{\tilde{s}}$, $\alpha$, $\theta$ and $\kappa$.  In light of recent LHC data we set $m_{\tilde{h}} = 125.5$ GeV and consider the mass range $120 < m_{\tilde{s}} < 600$ GeV.  As argued above, to a good approximation, the only search channel sensitive to the parameter $\alpha$ is for the decays $h \rightarrow \gamma \gamma$.

One could also consider coupling the scalars to additional colored fields, which would lead to enhanced production in the gluon fusion channel.  We will not consider this scenario here for two reasons.  Since the enhancement of the diphoton channel, and suppression of the non-diphoton channels, can be easily accommodated in the Higgs accomplice scenario there is little to gain by boosting the Higgs production in this way and the introduction of this additional parameter will not lead to a significant improvement in fitting the data.  Also, all Higgs production cross-sections in the models considered here are re-scaled in the same way, so above $150$ GeV, where the diphoton searches are not sensitive, one can employ the reported fits for all sub-channels combined.  Whereas to consider boosting gluon fusion alone means that all production sub-channels should be treated independently and the relative contributions of gluon fusion and vector boson fusion should be considered independently.  As these relative contributions and subsequent likelihoods must be estimated somehow, this leads to the introduction of further error.\footnote{For a treatment of this scenario with the friend decoupled see \cite{Riva,Falk}.}

\section{Statistical Methodology}
\label{sec:stats}

\subsection{Likelihood estimation}
Before considering scenarios with multiple Higgs-like scalars we briefly review some methodology regarding Higgs signal strength likelihood functions, which determine the compatibility of Higgs-like signal with the Higgs search results.\footnote{Throughout we employ a frequentist approach.  Employing a Bayesian approach allows the likelihoods for strength modifiers to be turned into probability density functions when one includes priors.  In this case, if strength modifiers depend on additional parameters then a Jacobian must be used when changing variables, so that the mode of the probability density function may not occur at the same place as the maximum of the likelihood  (see for example \cite{Cowan:1998ji}). 
The non-invariance of Bayesian estimators under reparametrization motivates our choice of a frequentist approach.}  Within a particular search channel, limits on a single Higgs particle of mass $m_h$ are expressed in terms of the strength modifier $\mu$, which relates the signal strength to that of a SM Higgs at a given mass, $\mu= n_s/(n_s)^{SM}$, where $n_s$ is the number of signal events expected for the particular search channel.
Given the number of observed events $n_{obs}$, one can construct the likelihood function $\mathcal{L}(n_{obs}|\mu,\boldsymbol{\theta})$ which is a function of the parameters $\mu$ and $\boldsymbol{\theta}$.  Here $\boldsymbol{\theta}$ stands for a set of nuisance parameters, which are fitted from the data to account for systematic effects and unknown background estimation parameters.

The standard quantity used to test hypotheses or set limits on $\mu$ is the so-called profile likelihood ratio \cite{Cowan:2010js}
\be
\lambda(\mu)=\frac{\mathcal{L}(n_{obs}|\mu,\doublehat{\boldsymbol{\theta}})}{\mathcal{L}(n_{obs}|\hat{\mu},\hat{\boldsymbol{\theta}})}~~,
\ee 
where $\doublehat{\boldsymbol{\theta}}$ is the value of $\boldsymbol{\theta}$ that maximizes the likelihood for a specified value of $\mu$,
while $\hat{\mu}$ and $\hat{\boldsymbol{\theta}}$ are the maximum likelihood estimators for $\mu$ and $\boldsymbol{\theta}$ respectively.
We do not have access to the full likelihood functions $\mathcal{L}_i(n_{obs}|\mu,\boldsymbol{\theta}_i)$ for the different channels, but given the information available from the experimental collaborations we can reconstruct approximate profile likelihood ratios $\lambda_i(\mu)$. In order to combine results from multiple search channels and different experiments, one should in principle calculate the profile likelihood ratio from the combined likelihood, which can be taken as the product of the different likelihoods if the channels are independent. Given that we do not know the full likelihood functions, we will take the combined profile likelihood ratio 
as the product of the $\lambda_i(\mu)$ reconstructed from single channels or experiments
\be
\lambda_C(\mu)\simeq\prod_i\lambda_i(\mu)~~.
\ee
It should be noted, however, that this approximation introduces an additional source of error if significant correlations between channels arise, possibly through the nuisance parameters.

In an abuse of terminology, we will henceforth refer to $\lambda(\mu)$ as the likelihood for $\mu$.   In order to reconstruct $\lambda_i(\mu)$, we note that, as described in \cite{wald,Cowan:2010js,Azatov:2012bz}, in the limit where the number of events is sufficiently large, with $n_{obs} \gtrsim 10$, the likelihood for a given channel can be approximated by
\be
\lambda(\mu)\simeq \exp^{-(\mu-\hat{\mu})^2 /2 \sigma^2_{obs}}  ~~,
\ee
where $\sigma_{obs}$ is in general a function of $\mu$.  For the $7$ TeV run the best fit strength modifier $\hat{\mu}$ and the error 
$\sigma_{obs}$ in individual $\gamma \gamma, \tau \tau, bb, WW$ and $ZZ$ search channels are reported by 
the ATLAS and CMS collaborations \cite{ATLAS-CONF-2012-019,CMS-PAS-HIG-12-008} as a function of the Higgs mass.
For most channels the Gaussian approximation, which assumes that $\sigma_{obs}$ is independent of $\mu$,
works well and using the best fit parameters and uncertainties the $95\%$ confidence limits can be reproduced well.
However in the $ZZ \rightarrow 4 l$ channel the likelihood function is clearly not Gaussian, as can be seen 
from the asymmetric confidence contours in \cite{ATLAS-CONF-2012-019,CMS-PAS-HIG-12-008}.  Thus the Gaussian assumption is not valid and its use can introduce 
artificial bias into parameter fits. 

We choose to approximate $\lambda(\mu)$ as a two-sided Gaussian, 
since this captures the approximately Gaussian nature of the likelihood and employs the three pieces of information available
at a given mass, namely the best fit point and two values of the log-likelihood away from the best fit point. 
Using the $7$ TeV data we can test this approximation by taking $\sigma$ on either side of the best fit value 
of $\mu$ from the $1 \sigma$ values (CMS \cite{CMS-PAS-HIG-12-008}), 
or $\Delta(-2\log\lambda)=1$ contours (ATLAS \cite{ATLAS-CONF-2012-019}) 
provided for the individual channels.  Although this approximation is crude, using the $7$ TeV data we find 
that when combining all channels and comparing with the reported combined best fit values the two-sided Gaussian 
assumption fares reasonably well, and typically better than the standard Gaussian approximation with symmetrized errors. 
It should be kept in mind that errors of $\mathcal{O} (10-15 \%)$ are typical using this approach,
combined with an inherent error due to digitization of the data, which we estimate to be as large as $\mathcal{O} (10 \%)$.

In some cases $\mu$ is the only free parameter, however in more complicated models involving modified Higgs couplings or additional
invisible decay widths $\mu$ becomes a function (in general different for different channels) of the additional parameters of the model, $\mu^i(\boldsymbol{\omega})$ where $\boldsymbol{\omega}$ denotes all the additional parameters and the superscript denotes the particular search channel. 
One can find best fit parameters by  maximizing the likelihood function and, since the quantity $-2 \log \lambda$ should approximately follow a chi-squared distribution \cite{Cowan:2010js}, one can also test the hypotheses of different models or signal strengths.

\subsection{Likelihood for multiple scalars}
Thus far we have only been concerned with models in which the hypothesis is of a single Higgs particle.  However in this work we consider models containing two Higgs-like scalars of mass $m_{\tilde{h}}$ and $m_{\tilde{s}}$, and we must estimate a combined likelihood for both.

The individual searches have differing mass resolutions, from as small as $1-3 \%$ in the $h\rightarrow \gamma \gamma$ and $h \rightarrow ZZ \rightarrow 4 l$ channels up to $20 \%$ in the $h \rightarrow \tau \tau$ and $h \rightarrow WW$ channels.  Whenever masses are greatly separated, i.e.\ $|m_{\tilde{h}} - m_{\tilde{s}}| \gg \sigma_i$ the hypothesized signal from one does not contaminate the search for the other, making the searches effectively independent.  In this case the likelihood can be taken as the product of the two independent likelihoods\footnote{The parameter dependence of the individual likelihoods is not necessarily independent.  In the case considered here one might wish, for example, to increase the mixing, which increases the signal from $\tilde{s}$ to explain some excess at $m_{\tilde{s}} \gg m_{\tilde{h}}$.  However doing so decreases the signal from $\tilde{h}$, which may be disfavored by the likelihood for $\tilde{h}$.} 
\be
\lambda_C\left[\mu_{\tilde{h}},\mu_{\tilde{s}},\boldsymbol{\omega},m_{\tilde{h}},m_{\tilde{s}}\right] = \lambda_{C}\left[\mu_{\tilde{h}},\boldsymbol{\omega},m_{\tilde{h}}\right] \times \lambda_{C}\left[\mu_{\tilde{s}},\boldsymbol{\omega},m_{\tilde{s}},\right]  ~~.
\ee
Whenever the Higgs and friend or accomplice are separated by mass splittings within or close to the mass resolution of a given search channel, $|m_{\tilde{h}} - m_{\tilde{s}}| \lesssim \sigma_i$, the situation becomes more complicated.  Without performing a full simulation we can still estimate the likelihood in such a scenario based on the mass resolutions provided.  However, in a conservative approach, we will not use this estimate to make precise statements about fits in regions where the signal from both scalars overlap, but will instead use it to determine the mass range in which the factorized likelihoods for the scalars can be trusted.

To understand how we estimate the combined likelihood whenever $|m_{\tilde{h}} - m_{\tilde{s}}| \lesssim \sigma_i$, one can first focus on a single search channel and consider a hypothetical situation in which signal from a SM Higgs, of mass $m_h$, is present in the data.  Performing a search for a Higgs of mass $m_h$, with cuts optimized for this mass, one expects to observe a certain number of signal events $n_s (m_h)$, and to reconstruct a strength modifier at that mass of $\mu(m_h) \approx 1$, up to statistical and systematic errors.  However, due to the finite mass resolution, a certain number of events, originating from the Higgs of mass $m_h$, may also pass the cuts for a Higgs search for a different mass $m'_h$.  Hence looking at searches for different masses one expects to observe a certain number of events $n_s (m'_h) < n_s (m_h)$ and to reconstruct a strength modifier at that mass $\mu(m'_h) < \mu(m_h)$, even though the true Higgs mass is $m_h$.  This makes intuitive sense: for a SM Higgs at mass $m_h$, with a finite amount of data one would not expect the reconstructed strength modifier to be a precise delta-function but rather it should follow some distribution which is peaked at $m_h$.

Given that we know $\mu \propto (n_{obs}-n_b) = n_s$ then, regardless of the mass-dependence of the backgrounds, we need only know the dependence of the eventual signal on the Higgs mass if we want to reconstruct the mass dependence of $\mu (m_h)$.  We choose to approximate the functional form to be Gaussian, such that if a SM Higgs is present at mass $m_h$ the number of signal events observed when applying the cuts, and hence searching, for a Higgs of mass $m'_h$, is
\be
n_s (m'_h) = n_s (m_h) \exp^\frac{-(m'_h-m_h)^2}{2 \sigma^2}  ~~,
\ee
where $\sigma$ is the mass resolution for that search.  For the case of multiple scalars the ATLAS $h \rightarrow WW$ search has been studied in \cite{Gupta:2011gd}.  The mass-dependence of the signal after cuts has been calculated and is shown in Fig 2.\ of \cite{Gupta:2011gd} for hypothetical Higgs-like scalars of mass $125$ and $170$ GeV.  One can see that due to the nature of the search the signal is not a delta-function centered at the Higgs mass but is rather a smooth distribution peaked at the true mass.  We find that a Gaussian provides a good fit to the data, and so we assume that $\mu (m_h)$ follows the same functional dependence.

Since the signal strength modifier is, by construction, normalized such that if a SM Higgs of mass $m_h$ is present in the data the strength modifier must be $\mu(m_h)\approx1$, we normalize the Gaussian distribution to have a peak value of $1$.  Given this assumption, combined with the approximate experimental resolution of the search channel, $\sigma_{i} (m)$, we estimate the strength modifier contributed by a SM Higgs of mass $m_h$ in a particular search channel to be
\be
\mu^i (m) = \exp^\frac{-(m-m_h)^2}{2 {\sigma_i}^2 ({m_h})} ~~,
\ee
where the normalization is chosen for a SM Higgs.  Clearly, to extend this to a non-SM Higgs one includes dependence on any additional parameters by rescaling production cross-sections and branching ratios accordingly.  Now to construct a likelihood for two Higgs scalars of mass $m_{\tilde{h}}$ and $m_{\tilde{s}}$ we estimate overlap of strength modifiers through the combination
\be
\mu^i (\boldsymbol{\omega},m_{\tilde{h}}) = \mu^i_{\tilde{h}} (\boldsymbol{\omega}) + \mu^i_{\tilde{s}} (\boldsymbol{\omega}) e^{-\frac{(m_{\tilde{s}}-m_{\tilde{h}})^2}{2 {\sigma_i}^2 (m_{\tilde{h}})}}  ~~~, ~~~~~~
\mu^i (\boldsymbol{\omega},m_{\tilde{s}}) = \mu^i_{\tilde{s}} (\boldsymbol{\omega}) + \mu^i_{\tilde{h}} (\boldsymbol{\omega}) e^{-\frac{(m_{\tilde{h}}-m_{\tilde{s}})^2}{2 {\sigma_i}^2 (m_{\tilde{s}})}}  ~~.
\label{eq:overlap}
\ee
In this way, if the mass splittings far exceed the experimental resolution the strength modifiers become independent and the likelihood factorizes into individual likelihoods for the independent scalars.  However as the masses approach one another signal overlap becomes important, and in the limit where the masses are equal the strength modifiers simply add together, as expected.  Alternatively one can think of this as the signal from one scalar acting as known background in the search for the other.  This method is clearly approximate, however it should give a reasonable estimate of the combined likelihood given the available information and is useful to determine the mass range in which the factorized likelihoods can be trusted.

We are combining multiple channels and so we must use different resolutions for each channel.  CMS reports the approximate mass resolution of the individual channels in \cite{Chatrchyan:2012tx}, which we use, taking the maximum value whenever a range is quoted.  We use $3\%$ for the $ZZ$ channel since this is the largest resolution in the individual $ZZ$ sub-channels which are sensitive to a light Higgs.  Our results will not be sensitive to this choice, since the dominant source of signal overlap is in the low-resolution channels unless $m_{\tilde{s}} \approx m_{\tilde{h}}$.  For ATLAS some resolutions are reported in \cite{ATLAS-CONF-2012-019} which are similar to those for the CMS searches.  When not reported, we assume the same resolution as in the CMS searches.  This assumption does introduce additional error whenever considering limits on scenarios where the scalars are close in mass, however in this region the dominant ATLAS sensitivity is in the $h \rightarrow \gamma \gamma$ and $h \rightarrow ZZ$ channels, with published resolutions, and the $h \rightarrow WW$ channel for which, by comparing with the results of \cite{Gupta:2011gd}, the assumption of $20 \%$ mass resolution is valid.  As a result it is likely that the overall error introduced into the combined limits and fits through this assumption is subdominant to other sources of error.  The chosen mass resolutions are detailed in \Tab{table:res}.

\begin{table}[h]
\caption{Approximate Light Higgs Search Mass Resolutions}
\centering
\begin{tabular}{c | c}
Channel ~~ & ~~ Resolution [$\%$]  \\ [0.5ex] 
\hline
$\gamma \gamma$ & 3 \\
$ZZ$ & 3 \\
$b b$ & 10\\
$\tau \tau$ & 20  \\
$WW$ & 20  \\ [1ex]
\end{tabular}
\label{table:res}
\end{table}

\subsection{Estimating the Importance of Signal Overlap}
\label{sec:estmate}
To estimate the impact of signal overlap on best fit parameters we consider the case with pure Higgs-singlet mixing, setting $\alpha=0$.  In this case strength modifiers for different search channels re-scale in the same way, simplifying the analysis.  To perform this estimate we only use the $7$ TeV data since best fit parameters and confidence contours are available for both CMS and ATLAS for the $h \rightarrow bb, \tau \tau, WW, ZZ$ and $h \rightarrow \gamma \gamma$ search channels.

\begin{figure}[h]
\centering
\includegraphics[height=2.9in]{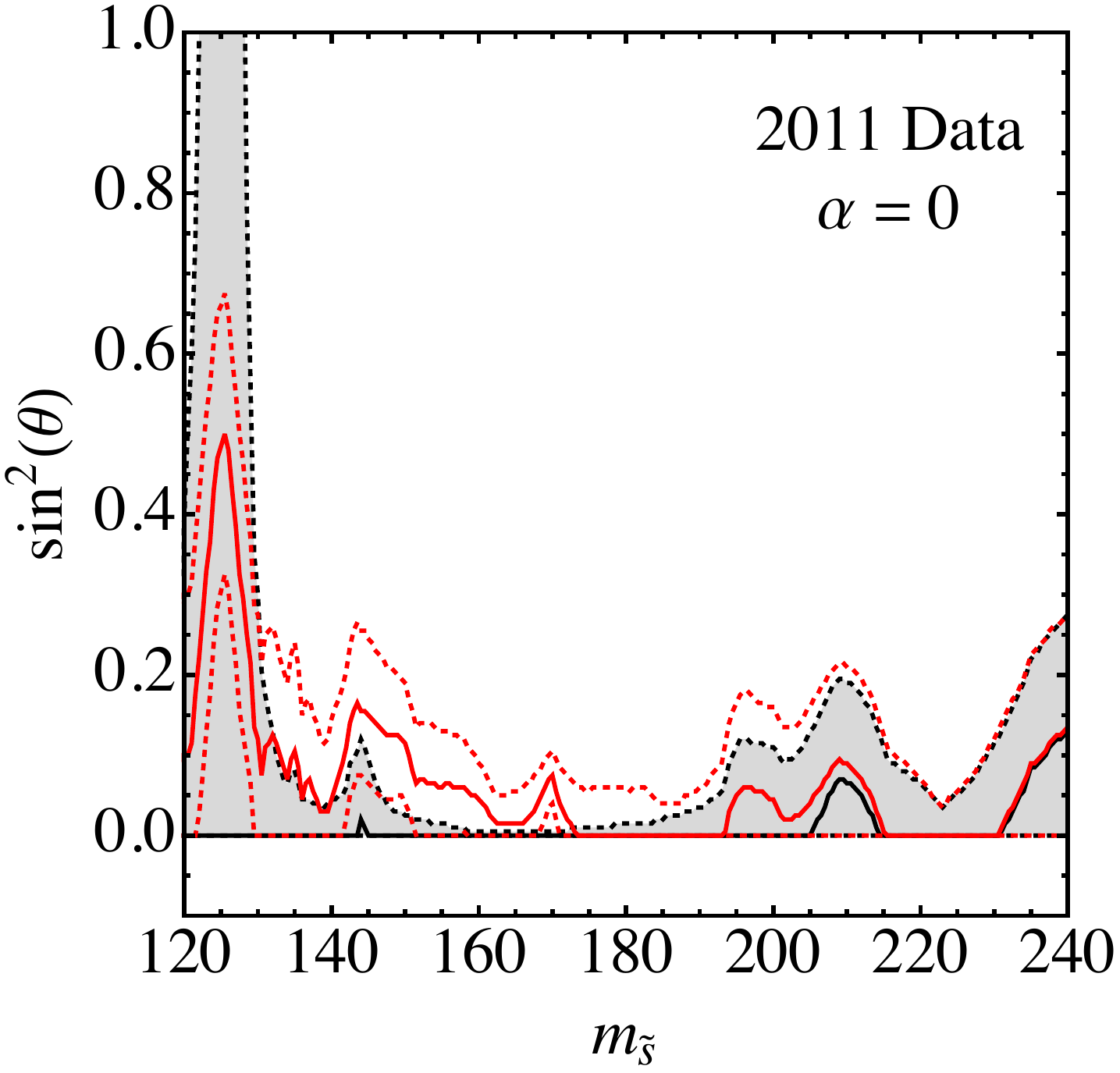}   
\caption{The best fit mixing angle as a function of the Higgs friend mass, $m_{\tilde{s}}$, for the combined likelihood with signal overlap included (black) and omitted (red).  $95\%$ confidence bands are also shown.  Above $m_{\tilde{s}} \sim210$ GeV the difference between both methods becomes negligible, demonstrating that above this mass the simple product of individual likelihoods can be trusted.  Below this mass the overlap of signal becomes important, suggesting that the simple individual likelihood products lose accuracy.}
\label{fig:alpha0low}
\end{figure}

In \Fig{fig:alpha0low} we plot, in black, the best fit mixing angle as a function of the singlet mass, $m_{\tilde{s}}$, for the combination of likelihoods of both scalars with signal overlap included according to \Eq{eq:overlap}.  Due to a deficit in background events a negative Higgs event rate is, at some masses, preferred by the data.  However, since we are fitting to a model restricted to real mixing angles, such negative event rates are not within the parameter space of the model, and these points usually correspond to a best-fit value of $\theta=0$.  We calculate $95\%$ confidence bands by finding the mixing angle at which $\Delta ( -2 \log\lambda) \equiv -2 \left(\log\lambda (\mu)-\log\lambda(\hat{\mu})\right) = 3.84$.  As argued in the caption, for $m_{\tilde{s}} \lesssim 210$ GeV the overlap in signal clearly becomes important, and the simple product of likelihoods should not be used.

However, since the best fit values shown in black do accommodate the signal overlap to some degree, we can still extract some qualitative features.  Typically for masses $m_{\tilde{s}}\lesssim 200$ GeV the preferred mixing angle is $\theta \approx0$.  This is due to two dominant effects.  First of all, small mixing angles are preferred for the fit of $m_{\tilde{h}}=125.5$ GeV to the $7$ TeV data since the signal at this mass prefers $\mu\sim 0.8$, and larger mixing angles which reduce the signal further are penalized.  Second, the strong limits for $m_{\tilde{s}}\lesssim200$ GeV also prefer the signal from the friend to be small, requiring a small mixing angle.  One can also see that for $m_{\tilde{s}}\approx m_{\tilde{h}}$ the mixing angle essentially becomes unconstrained.  This is due to the fact that in this case each strength modifier is almost independent of $\theta$ since $\mu \approx \sin^2(\theta)+\cos^2(\theta) \approx1$.

\section{Results}
\label{sec:results}

\subsection{Data}
As demonstrated in \Sec{sec:estmate}, for $m_{\tilde{s}} \gtrsim 210$ GeV we can effectively treat the likelihoods for the scalars individually, taking the product to find the combined likelihood.  In this mass range the strength modifiers for the friend in the relevant channels scale in the same way, since the diphoton search is not sensitive, and we only require the likelihood for the combined channels.  Combined $7$ and $8$ TeV best fit values and confidence contours for the ATLAS searches were presented in \cite{:2012gk} and so we employ these to construct the two-sided Gaussian likelihood.  Combined $7$ and $8$ TeV expected and observed $95\%$ confidence limits for this mass range were presented for CMS in \cite{CMS-PAS-HIG-12-020}, and so we use this data to estimate the combined likelihood in this region.  In \cite{Cowan:2010js} it is shown that the best fit strength modifier can be simply approximated by the difference of the observed and expected upper limits, and the error is given by $\sigma \approx \mu^{95\%}_{exp}/1.64$.

For the Higgs likelihood at $125.5$ GeV we require information on individual channels since the parameter dependence differs for the $\gamma \gamma$ channel from the other channels.  For this low mass region in \cite{:2012gk} best fit values and confidence contours for the $\gamma \gamma$, $ZZ$ and $WW$ channels for combined $7$ and $8$ TeV runs at ATLAS are given.  We estimate the full ATLAS likelihood function using these three channels combined with $7$ TeV values for the $b b$ and $\tau \tau$ channels \cite{ATLAS-CONF-2012-019}, all at $125.5$ GeV.  For CMS best fit values and uncertainties for the combined $7$ and $8$ TeV runs for all channels have been presented for a Higgs at $125.5$ GeV \cite{:2012gu}, and we use this to estimate the CMS likelihood function.  Since these fits have been presented for a $125.5$ GeV Higgs we assume this Higgs mass throughout.  We also use the recent Tevatron data \cite{:2012zzl}, taking the best fit values and uncertainties for the $\gamma \gamma$, $WW$ and $bb$ channels.

With this information we can estimate the full likelihood for both scalars to determine whether a social Higgs allows for any improvement in fitting the data when compared with a SM (antisocial) Higgs.

\subsection{Higgs Friend Scenario}

\begin{figure}[h]
\centering
\includegraphics[height=2.7in]{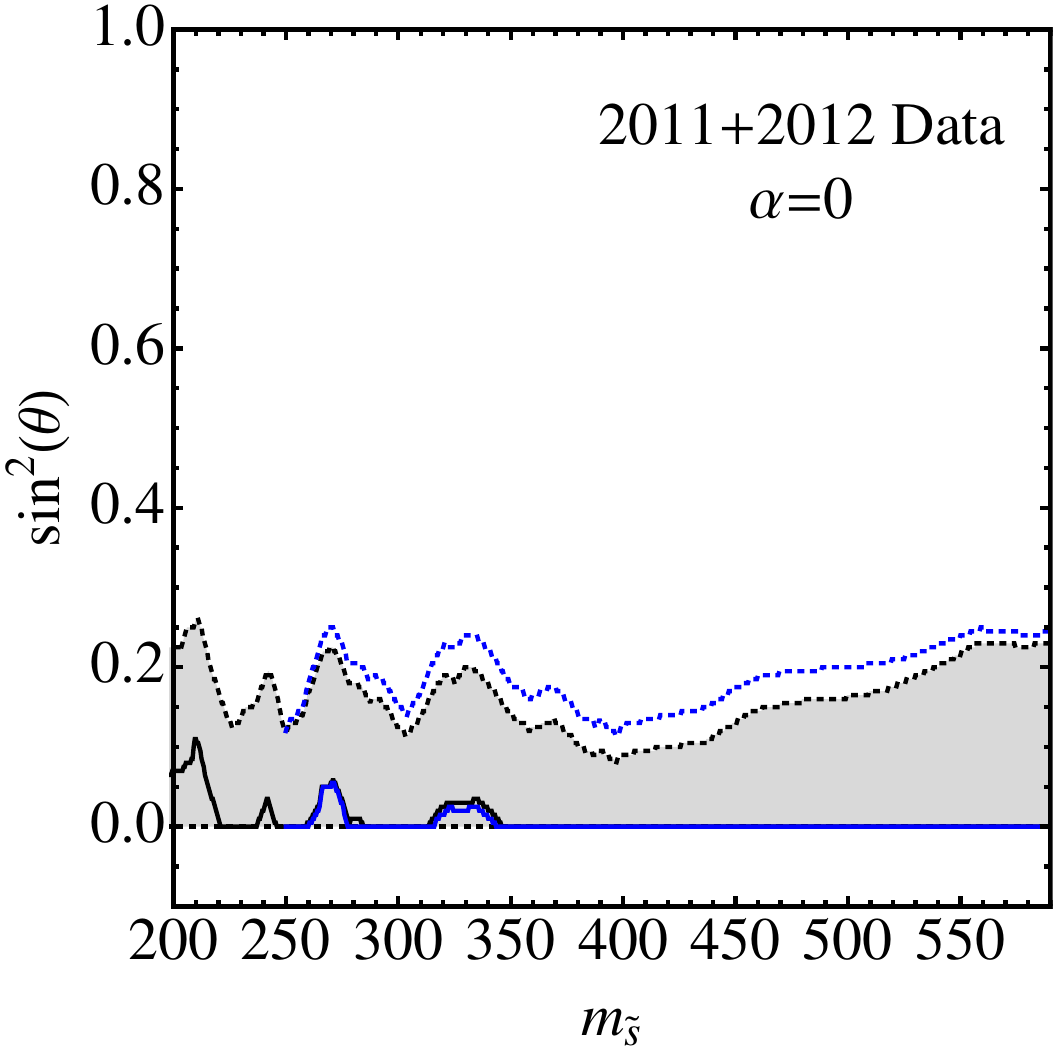}     \hspace{0.4in}  \includegraphics[height=2.7in]{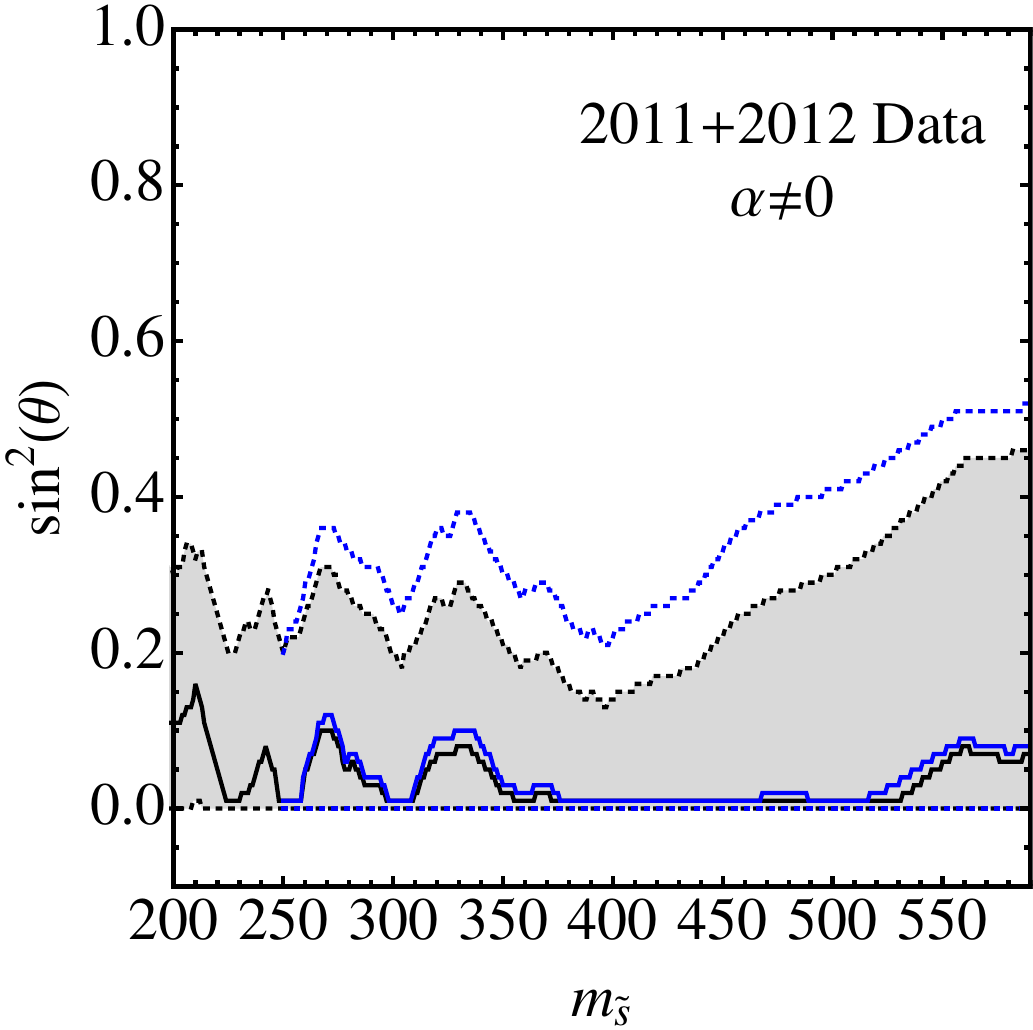}   
\caption{The best fit mixing angle as a function of the scalar mass, $m_{\tilde{s}}$, at high masses.  The limit of vanishing $\tilde{s}\rightarrow 2 \tilde{h}$ branching ratio is shown in black and for $\kappa = 0.5$ (see \Eq{eq:BR}) in blue.  $95\%$ confidence contours are also shown (dashed) in corresponding colors.  On the left panel we set $\alpha=0$ and allow only mixing with the Higgs friend, with $95\%$ confidence contours determined via $\Delta (-2 \log\lambda) = 3.84$.  On the right panel we allow for enhanced $\tilde{h} \rightarrow \gamma \gamma$ decays in the Higgs accomplice scenario and find the best fit values of $\theta$ and $\alpha$.  In this case we find $95\%$ confidence contours by finding the maximum value of $\theta$ for which $\Delta (-2 \log\lambda) = 5.99$.  On both panels it is clear that due to strong limits in the region $380$ GeV $\lesssim m_{\tilde{s}}\lesssim 450$ GeV the SM is preferred over both scenarios.  However, due to mild excesses or weak limits at other masses both scenarios can slightly improve the fit to the data in comparison to the SM.  Due to a small excess in $h\rightarrow ZZ$ events in ATLAS and CMS, around $m_{\tilde{s}} \approx 210$ GeV the accomplice scenario becomes marginally preferred over the SM at $95\%$, while for the friend scenario the SM lies on the $95\%$ confidence contour.  Following the discussion of \Sec{sec:estmate} the reader should keep in mind that in the region below $m_{\tilde{s}}\approx 210$ GeV some error is introduced by neglecting signal overlap.}
\label{fig:alphahigh}
\end{figure}

First we consider the friend scenario, where the Higgs is mixed with a singlet scalar.  On the left-hand panel of \Fig{fig:alphahigh} we plot the best fit mixing angle as a function of the singlet mass for $m_{\tilde{s}}>200$ GeV.  The best fit for the limit of vanishing invisible width is shown in black.  We also plot results allowing for the decays $BR (\tilde{s} \rightarrow 2 \tilde{h})$ following \Eq{eq:BR} with $\kappa =0.5$ in blue.  Constraints are weakened by this effectively invisible width due to suppression of the signal, and fits previously requiring some mixing now require greater mixing due to dilution of the signal at high masses.

On the whole, since the best fit points satisfy $\sin^2 (\theta) \lesssim 0.1$ and are consistent at $95\%$ with $\sin^2 (\theta) =0$ it is clear that the SM provides almost as good a fit as the mixed model, and for the majority of the mass range the Higgs friend scenario provides no advantage over the SM, even though some mixing is preferred for the $125.5$ GeV signal.  This is not true, however, at one point near $m_{\tilde{s}} \sim 210$ GeV, where the Higgs friend scenario provides some improvement in fit over the SM, and the SM ($\theta=0$) actually lies close to the $95\%$ confidence contour.  This improvement in fit is not great enough to suggest strong evidence for the presence of a Higgs friend, but is interesting nonetheless.   If considered for just the singlet, this excess prefers $\mu \sim 0.2$, and the Higgs signal at $125.5$ GeV prefers $\mu \sim 1.1$ and so both excesses fit the Higgs friend model well for $\sin^2 (\theta) \sim 0.1$.  Small excesses or weak limits at other masses also allow for a best fit with non-zero $\theta$, although $\theta=0$ lies within the $95\%$ confidence contour.  

Also, it is interesting to note that in some cases mixing angles as small as $\sin^2 \theta \sim 0.1$ are disfavored at the $95\%$ level, showing that the Higgs searches are in some cases already sensitive to relatively minor modifications of the Higgs sector.  We can also consider the case where the friend is effectively decoupled, with mass beyond the current sensitivity of the LHC searches.   In this case we find that $\sin^2 (\theta) = 0$ is preferred, with an error of $0.09$, consistent with the SM.

\subsection{Higgs Accomplice Scenario}
Whenever we allow for enhancement of the decays $\tilde{h} \rightarrow \gamma \gamma$ by coupling the friend to photons then, regardless of the value of $\theta$, we can always choose the coupling, $\alpha$, such that $\mu_{\gamma \gamma}>1$ can be reproduced for the Higgs signals at $125.5$ GeV.  The other search channels only constrain $\theta$.  Furthermore, as the diphoton searches look for resonances below $150$ GeV, for $m_{\tilde{s}} > 150$ GeV the likelihood function for $\tilde{s}$ is independent of $\alpha$.  Hence $\alpha$ allows the freedom to achieve the desired Higgs diphoton rates without degrading the fit to the Higgs friend.  For the $125.5$ GeV excess there is a tantalizing hint that the $\gamma \gamma$ channel might be enhanced, while other rates might be suppressed, perhaps suggesting non-zero mixing angles.  As such the Higgs accomplice scenario allows for a better fit to the data than the simple Higgs friend scenario. 

On the right-hand panel of \Fig{fig:alphahigh} we show the best fit values for $\theta$ as a function of $m_{\tilde{s}}$.  The $95\%$ confidence contours are found by finding the largest possible value of $\theta$ such that $\Delta ( -2 \log\lambda) =5.99$.  Thus a best fit value of $\alpha$, which is typically an $\mathcal{O}(1)$ number (when positive), is concealed within this plot.

Over the mass range $\sin^2 (\theta) \lesssim 0.2$ is preferred, with a slight increase in the best fit values due to the ability to accommodate the slightly suppressed rates in the non-diphoton channels at $125.5$ GeV without suppressing the diphoton rates.  At high masses weaker limits allow for larger signal, and hence mixing angles, to be accommodated for the Higgs accomplice while simultaneously fitting the slightly suppressed channels for the Higgs.  The best fit value near $210$ GeV has also increased slightly in comparison with the friend scenario, as expected, and a slight bump in the lower $95\%$ confidence limit shows that, at $95\%$, the SM is disfavored in comparison to the Higgs accomplice scenario whenever $m_{\tilde{h}} = 125.5$ GeV and $m_{\tilde{s}} = 210$ GeV.  Thus the Higgs accomplice scenario accommodates a better fit due to the additional source of $\gamma \gamma$ decays, however, the preference for this scenario is still not particularly strong.

\begin{figure}[h]
\centering
\includegraphics[height=2.5in]{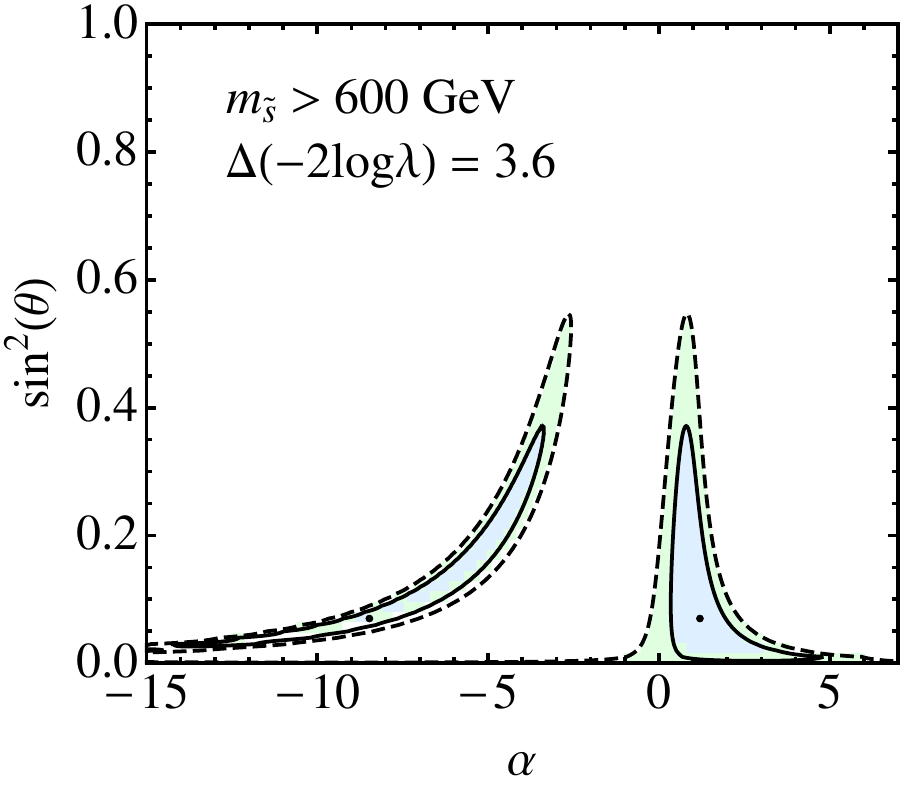}   \hspace{0.2in} \includegraphics[height=2.5in]{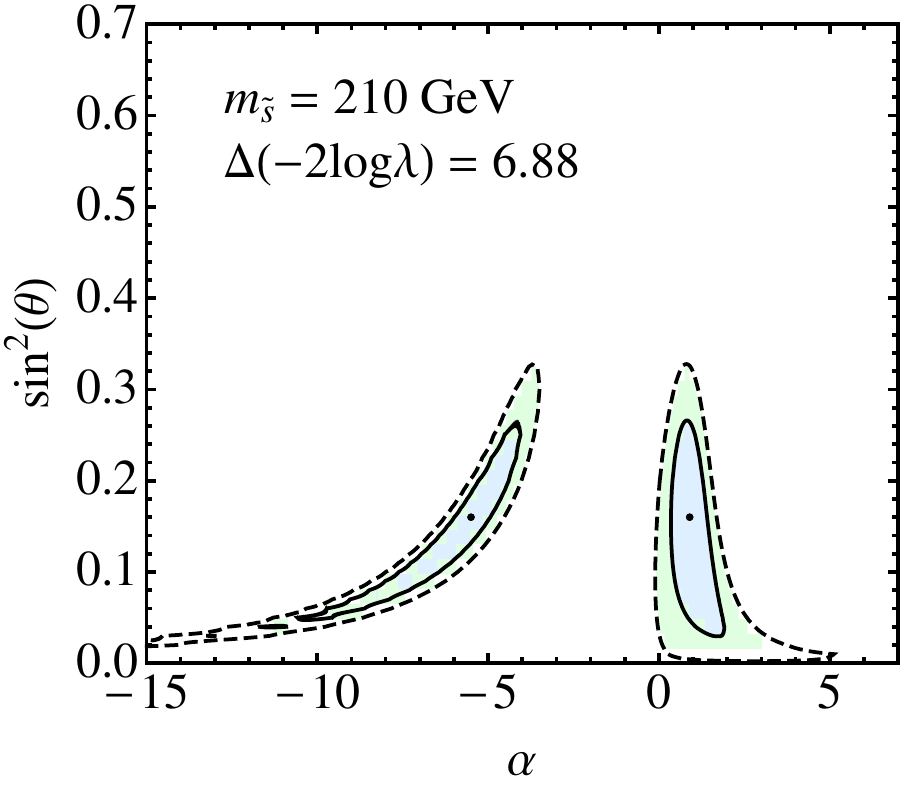} 
\caption{Best fit points and $68\%$ and $95\%$ confidence contours, corresponding to $\Delta (-2 \log\lambda) = 2.28, 5.99$ for the specific scenario of a Higgs accomplice beyond collider reach (left panel) and a Higgs accomplice at $210$ GeV (right panel).  The left panel shows the fit for the Higgs to solely the $125.5$ GeV data since the accomplice is decoupled.  Due to enhancement of $\tilde{h} \rightarrow \gamma \gamma$ and suppression of other channels, non-zero mixing angles are preferred, alongside $\mathcal{O} (1)$ values of $\alpha$.  The SM, $\sin^2(\theta)=0$, is within the $95\%$ confidence contour.  The change in $-2 \log\lambda$, which can be interpreted as the change in $\chi^2$, from the best fit point to the SM is also shown, and is marginally greater than the number of extra parameters introduced.  When signal from the Higgs accomplice is included at $210$ GeV, and the likelihoods for both scalars are combined (right panel), the overall fit is improved significantly and the SM becomes marginally disfavored at greater than $95\%$.}
\label{fig:alphathetafit}
\end{figure}

In \Fig{fig:alphathetafit} we show the best fit values as well as $68\%$ and $95\%$ confidence contours, corresponding to $\Delta ( -2 \log\lambda) =2.28$ and  $\Delta( -2 \log\lambda) =5.99$, for $\alpha$ and $\sin^2(\theta)$ whenever $\tilde{s}$ is decoupled and doesn't contribute any signal in the search window (left panel) and whenever signal from the friend is present at $m_{\tilde{s}} = 210$ GeV (right panel).   The left panel shows that with the friend decoupled the best fit points prefer non-zero mixing, since the $\gamma \gamma$ rates can be fit independently of $\theta$.  The SM at $\theta=0$ is within the $95\%$ confidence contour, showing there is no strong preference for this scenario.

Comparing both panels of \Fig{fig:alphathetafit}, we see that the suppression of the non-diphoton event rates at $125.5$ GeV requires smaller mixing angles than are required to fit the small excess at $210$ GeV, and the best fit mixing angle moves to larger values when the friend is included.  On the right-hand panel the SM, $\theta=0$, lies outside the $95\%$ confidence contour, demonstrating that if the accomplice mass is close to $210$ GeV the Higgs accomplice scenario gives a definite improvement over the SM in fitting the data.  In particular, a change of $\Delta (-2 \log\lambda) \sim 7$ between the SM and the best fit point indicates that the improvement in fit is not negligible, since only three new parameters are introduced (if one considers the accomplice mass to be fixed).\footnote{One might worry that the model is effectively over-fitting the data.  Combining the measurements for the $125.5$ GeV and $210$ GeV Higgs searches one finds a log-likelihood value (which can be considered as approximately representing the $\chi^2$ value) of $-2 \log {\lambda} = 15$ for the SM with $D = 13+2-2=13$ degrees of freedom ($125.5$ GeV channels, $210$ GeV combined channels, and $m_h, \mu$). For the Higgs accomplice model, at the best fit parameters shown on the right-hand panel of \Fig{fig:alphathetafit}, we find $-2 \log {\lambda} = 8$ for $D = 13+2-2-3=10$ degrees of freedom.  Since $\chi^2/D \sim 1$ for this model the data is not being over-fitted, and much of the improvement compared to the SM comes from fitting the excess at $210$ GeV, which can not be accommodated within the SM.}

Overall, the Higgs accomplice scenario does allow for a significant improvement in fitting the data over the SM, however this improvement is not great enough to claim support, or evidence, for a social Higgs.

\section{Electroweak Precision Constraints}
\label{sec:electroweak}
As pointed out in \cite{Bowen:2007ia}, if the Higgs mixes with a friend then precision electroweak observables are altered in comparison to the SM.  In particular, W and Z boson couplings to the Higgs are suppressed, and the friend can also enter at one loop into self-energy graphs.  Here we study the differences in the S and T parameters \cite{Peskin:1990zt,Peskin:1991sw} relative to the SM for the Higgs friend model.  We calculate these differences at one loop by taking the Higgs contributions to S and T from \cite{Hagiwara:1993ck,Hagiwara:1994pw} and re-scaling them by $\cos^2(\theta)$.  We also add a similar contribution for the friend and then subtract off values for a SM Higgs at $125.5$ GeV.

\begin{figure}[h]
\centering
\includegraphics[height=2.4in]{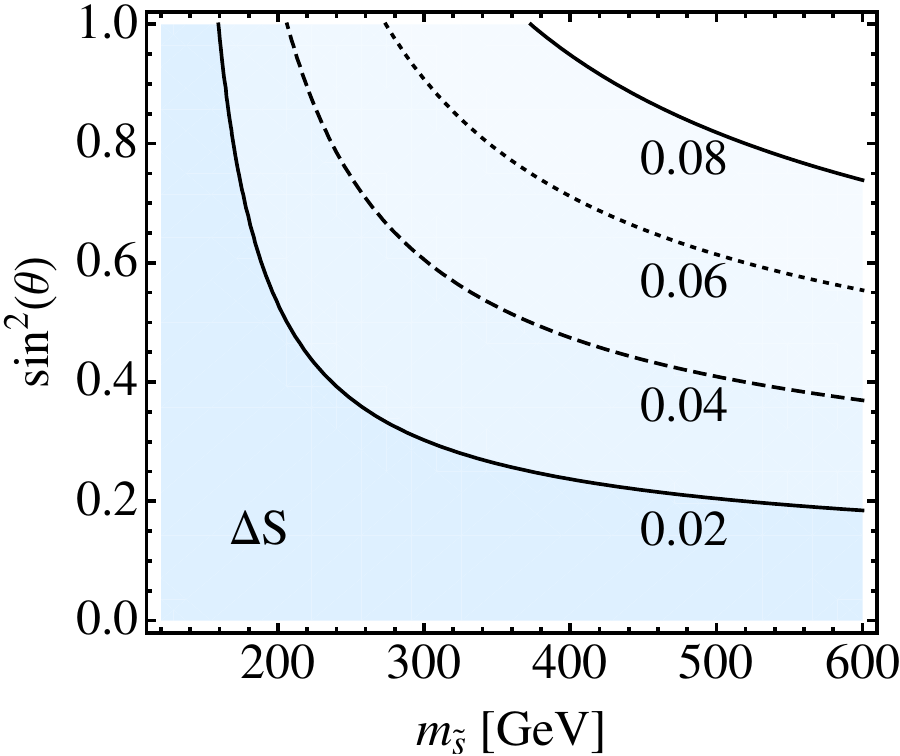}   \hspace{0.2in} \includegraphics[height=2.4in]{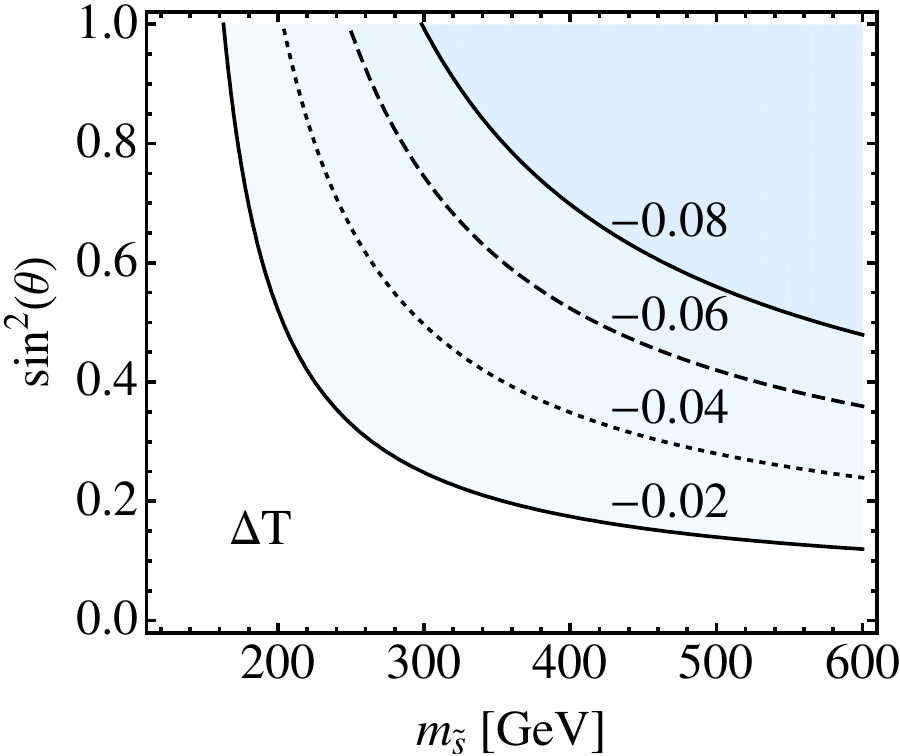}    
\caption{Contours of $\Delta S = S(\tilde{h},\tilde{s},\theta) - S (h)$ and $\Delta T = T(\tilde{h},\tilde{s},\theta) - T (h)$, for the simple Higgs friend model.  We have set $m_{\tilde{h}} = m_h = 125.5$ GeV.  For the majority of parameter space this model is consistent with electroweak precision data at $1 \sigma$.}
\label{fig:EW}
\end{figure}

In \Fig{fig:EW} we show contours of the change in the S and T parameters relative to the SM with a Higgs at $125.5$ GeV.  In \cite{PDG} for a Higgs mass in the range $115.5 < m_h < 127$ GeV the S and T parameters are given as $S = 0.00^{+0.11}_{-0.10}$ and $T = 0.02^{+0.11}_{-0.12}$, so in this model deviations from the SM are typically within $1 \sigma$, even with relatively large mixing angles, hence electroweak precision places no strong constraints on models with a relatively light Higgs friend.  Heavier Higgs friends are less consistent with electroweak precision constraints, however agreement at close to $1 \sigma$ can still be found for friends with masses greater than $1$ TeV \cite{Bowen:2007ia}.

Statements about precision electroweak observables are model-dependent and if additional electroweak-charged fields are present, as in the Higgs accomplice scenario, or as would be expected in a complete model which addresses fine-tuning issues, then further alterations to the S and T parameters would arise.  In either case one must then consult the particular model to establish consistency with electroweak precision data.  As such, the bounds shown here should be considered a demonstration of consistency in the friend scenario, rather than a reflection of the consistency of a possible underlying theory.

\section{Discussion}
\label{sec:discussion}
The resonance with a mass near $125$ GeV recently discovered at the CERN LHC exhibits properties consistent with the SM Higgs boson.  Only analyses of future data can convincingly determine whether or not it is indeed the SM Higgs, however, since there is currently no strong evidence to the contrary, it is now possible to constrain scenarios where the Higgs properties are significantly altered.  Furthermore, null results in Higgs searches at other masses already place strong bounds on neutral scalars with Higgs-like production and decay properties.  Motivated by this observation, in this work we have examined the impact this has on two simple models, the Higgs friend and Higgs accomplice scenarios, which may act as simplified models for theoretically motivated extended Higgs sectors, such as arise in the NMSSM.  Both scenarios are still compatible with the data, however large mixing angles $\theta \gtrsim \pi/4$ are typically disfavored at the $95\%$ level.  Small mixing angles satisfying $\sin^2 (\theta) \lesssim 0.2$, can improve the overall fit for the Higgs at $125.5$ GeV, especially if the model accommodates enhanced couplings to photons.  However, the improvement in fit is, in the majority of cases, not statistically significant. The only exception is for a Higgs accomplice with a mass near  $210$ GeV which allows for an improvement in fitting the data at greater than $95\%$ significance when compared to the SM .  However, keeping the numerous uncertainties and the low statistical significance of the excess in mind, one cannot interpret this as evidence for a bone-fide Higgs friend or accomplice.


\acknowledgments{We benefitted from conversations with Aleksandr Azatov, Adam Falkowski, Jamison Galloway and Eric Kuflik.  We are grateful to Jesse Thaler for useful comments and to John March-Russell for conversations, comments on an early draft of this paper, and for suggesting the title.  M.M. and D.B. are supported by the U.S. Department of Energy (DOE) under cooperative research agreement DE-FG02-05ER-41360.  M.M. is also supported by a Simons Postdoctoral Fellowship and D.B. is also supported by Istituto Nazionale di Fisica Nucleare (INFN) through a  ``Bruno Rossi" Fellowship.}

\bibliographystyle{JHEP}
\bibliography{HiggsFriendref}

\end{document}